\documentclass[ prl, twocolumn, nofootinbib, floatfix,superscriptaddress]{revtex4-2}

\usepackage{notes2bib}
\usepackage{natbib}
\usepackage{gensymb}
\usepackage{amsthm}
\usepackage{amsmath}
\usepackage{color}
\usepackage{bm}
\usepackage{amsmath,amssymb}
\usepackage{amsfonts}
\usepackage{dsfont}
\usepackage{graphicx}
\usepackage{float}
\usepackage{subfigure}
\usepackage{verbatim}
\usepackage{amsfonts}
\usepackage{bbold}
\usepackage{dcolumn}
\usepackage{bm}
\usepackage[dvipsnames]{xcolor}
\usepackage{listings}
\definecolor{blueprl}{RGB}{46,48,146}
\usepackage[pdftex,colorlinks=true,urlcolor=blueprl,citecolor=blueprl,linkcolor=blueprl]{hyperref}
\usepackage{times}
\usepackage{color}
\usepackage[dvipsnames]{xcolor}

\usepackage[mathscr]{euscript}
\usepackage{amsmath}
\usepackage{graphicx}
\usepackage{dcolumn}
\usepackage{bm}
\usepackage{epsfig}
\usepackage{amssymb,latexsym,mathrsfs}
\usepackage{graphicx}
\usepackage{color}
\usepackage{hyperref}
\usepackage{float}
\usepackage{diagbox}
\usepackage{ulem}

\definecolor{vividviolet}{rgb}{0.62, 0.0, 1.0}
\definecolor{amaranth}{rgb}{0.9, 0.17, 0.31}
\definecolor{palatinateblue}{rgb}{0.15, 0.23, 0.89}
\definecolor{brightpink}{rgb}{1.0, 0.0, 0.5}
\definecolor{cornflowerblue}{rgb}{0.39, 0.58, 0.93}
\definecolor{deepcarminepink}{rgb}{0.94, 0.19, 0.22}
\definecolor{radicalred}{rgb}{1.0, 0.21, 0.37}
\definecolor{blueblue}{RGB}{21,47,181}
\definecolor{greengreen}{RGB}{65,166,16}

\newcommand{\be}{\begin{equation}}
\newcommand{\ee}{\end{equation}}
\newcommand{\bs}{\begin{split}} 
\newcommand{\bea}{\begin{eqnarray}}
\newcommand{\eea}{\end{eqnarray}}
\newcommand{\non}{\nonumber }

\usepackage{mathtools}
\newcommand{\D}{\mathrm{d}}
\newsavebox{\myhbar}

\newcommand{\p}{\partial}

\begin{document}


\title{ Quantum signatures of black hole mass superpositions}
\author{Joshua Foo}
\email{joshua.foo@uqconnect.edu.au}
\affiliation{Centre for Quantum Computation \& Communication Technology, School of Mathematics \& Physics, The University of Queensland, St.~Lucia, Queensland, 4072, Australia}
\author{Cemile Senem Arabaci}
\affiliation{Department of Physics and Astronomy, University of Waterloo, Waterloo, Ontario, Canada, N2L 3G1}
\author{Magdalena Zych}
\affiliation{Centre for Engineered Quantum Systems, School of Mathematics and Physics, The University of Queensland, St. Lucia, Queensland, 4072, Australia}
\author{Robert B.\ Mann}
\affiliation{Perimeter Institute, 31 Caroline St., Waterloo, Ontario, N2L 2Y5, Canada}
\affiliation{Department of and Astronomy, University of Waterloo, Waterloo, Ontario, Canada, N2L 3G1}

\begin{abstract}
We present a new operational framework for studying ``superpositions of spacetimes'', which are of fundamental interest in the development of a theory of quantum gravity. Our approach capitalizes on nonlocal correlations in curved spacetime quantum field theory, allowing us to formulate a metric for spacetime superpositions as well as characterizing the coupling of particle detectors to a quantum field. We apply our approach to analyze the dynamics of a detector (using the Unruh-deWitt model) in a spacetime generated by a BTZ black hole in a superposition of masses. We find that the detector exhibits signatures of quantum-gravitational effects corroborating and extending Bekenstein's seminal conjecture concerning the quantized mass spectrum of black holes in quantum gravity. Crucially, this result follows directly from the approach, without any additional assumptions about the black hole mass properties.
\end{abstract} 

\date{\today} 

\maketitle 

\textit{Introduction}\textemdash Black holes continue to captivate physicists from a diverse array of backgrounds, from cosmology and astroparticle physics, to quantum field theory and general relativity. Because of the extreme gravitational environments they generate, they are considered primary candidates for studying regimes in which quantum gravity effects are present \cite{gibbons1993euclidean,wald1984black,nicolinidoi:10.1142/S0217751X09043353}. Indeed, the discoveries of Hawking radiation \cite{hawking1974black} and black hole evaporation \cite{hawkingcmp/1103899181}  gave rise to the well-known information paradox \cite{hawkingPhysRevD.14.2460,Giddings:1995gd,Mathur_2009,Harvey:1992xk} and an entire field seeking its resolution, which aptly illustrates the existing conflicts between quantum theory and general relativity. 

Bekenstein \cite{bekensteinPhysRevD.7.2333,bekenstein2020quantum,reggePhysRev.108.1063} was among the first to recognise that a complete theory of quantum gravity must account for the treatment of black holes as quantum objects. His key insight was in demonstrating that the black hole horizon area, and hence its mass, is an adiabatic invariant, with an associated discrete quantization \cite{hodPhysRevLett.81.4293} and evenly spaced energy levels \cite{BEKENSTEIN19957}. These pioneering studies have given rise to the burgeoning research field of quantum black holes \cite{Bekenstein:1997bt,BANERJEE2010279,Vagenas:2008yi,POURHASSAN2017325,LOCHAN201637,Kwon_2010}. 

The existence of mass-quantized black holes implies that they may also exist in superpositions of mass eigenstates. A mass-superposed black hole is an example of a quantum superposition of spacetimes; since the different masses individually define a unique classical solution to Einstein's field equations, the resulting amplitudes of the mass superposition correspond to associated ``spacetime states''. Understanding the effects that arise in such spacetime superpositions is an important stepping stone towards developing complete description 
of a quantized spacetime. ``Top-down'' approaches  -- those seeking  to formally quantize general relativity 
-- date back to the  Wheeler-deWitt (WdW) equation \cite{Wheeler1968SUPERSPACEAT,dewittPhysRev.160.1113}, a wave functional of the metric that offered a conceptual scheme for dealing with superpositions of spacetime geometries \cite{rovelli2015covariant,rovelli2008loop}. Building upon this work, canonical quantization techniques were applied to the metric variables \cite{kuchaPhysRevD.50.3961,THIEMANN1993211} leading to the development of loop quantization \cite{ashtekarPhysRevLett.57.2244,ashtekarPhysRevLett.80.904} which in turn, has yielded solutions including black holes in a superposition of masses \cite{rovelli2004quantum,KASTRUP1994665,Campiglia_2007,Gambini_2014,gambiniPhysRevLett.110.211301,Demers_1996,Kiefer_2013}.

Recently, there has been significant interest in the study of spacetime superpositions from an operational viewpoint, grounded in the definition of spacetime events using physical devices (clocks, rods, and detectors). Examples include investigations into the resulting quantum causal structures \cite{zych2019,Henderson:2020zax} quantum reference frames and the equivalence principle~\cite{Giacomini2021spacetimequantum,giacomini2021einsteins,giacomini2021quantum,belenchiaPhysRevD.98.126009}, and fundamental decoherence mechanisms \cite{Arrasmith_2019}, along with applications to analog gravity~\cite{barcelo2021superposing}, and tabletop experiments aimed at testing the quantum nature of gravity~\cite{christodoulou2019possibility,marletto2017gravitationally,bose2017spin}. A recent investigation developed an approach where a particle detector (modelled as the Unruh-deWitt detector~\cite{RevModPhys.80.787}) evolves in a de Sitter spacetime in superpositions of spatial translations and curvature, to study the quantum effects arising in such a background \cite{Foo_2021}. However, the conformal equivalence of de Sitter and Rindler spacetime (a uniformly accelerated reference frame in Minkowski spacetime), meant that even in such scenario some of the effects were  equivalent to those experienced by a detector in a superposition of semiclassical trajectories in Minkowski spacetime~\cite{fooPhysRevD.102.085013,barbadoPhysRevD.102.045002}.

We are thus motivated by a problem of fundamental interest, namely the detection of genuinely quantum-gravitational effects that do not arise for relativistic quantum matter in a classical spacetime~\cite{Zych:2015fka}.
In this Letter we propose an operational method for constructing and quantitatively analyzing effects produced by general spacetime superpositions, and apply it to the analytically tractable (2+1)-dimensional Banados-Teitelboim-Zanelli (BTZ) black hole. We show that the black hole mass superposition elicits novel effects with no direct analogue in a classical spacetime. We obtain a ``conditional''  metric for the spacetime superposition, which builds upon approaches in which correlations in quantum field theory act as a proxy for spacetime distance \cite{Saravani_2016,Kempf:2021xlw}. These (two-point) field correlations include those evaluated between the different spacetime amplitudes in superposition~\cite{fooPhysRevD.103.065013,fooPhysRevResearch.3.043056}. In this way we are able to calculate the response of a particle detector in the non-classical spacetime considered. As the detector interacts with the Hawking radiation of the black hole, its response experiences resonances at squared rational values of  ratio between the superposed mass values. 
This effect provides a novel, independent signature that  supports and extends Bekenstein's conjecture regarding the discrete mass eigenspectrum of quantum black holes.
Throughout  we utilize natural units, $\hslash = k_B = c = 8G = 1$.

\textit{The BTZ black hole}\textemdash The BTZ black hole is a (2+1)-dimensional solution to Einstein's field equations with a negative cosmological constant   $\Lambda = - {1/l^2}$ where $l$ is the anti-de Sitter (AdS) length scale (the BTZ spacetime is asymptotically AdS) \cite{banadosPhysRevLett.69.1849,Carlip:1995qv,carlip2003quantum}.  While our method of coupling matter to a superposition of spacetimes applies to any metric and spacetime dimension,
the advantage of studying 
the BTZ spacetime first is that 
its field correlation functions have analytic closed forms \cite{lifschitzPhysRevD.49.1929,hodgkinsonPhysRevD.86.064031,Smith_2014,Henderson_2018,robbins2020entanglement,HENDERSON2020135732,CAMPOS2021136198,robbins2021antihawking}, whereas in its (3+1)-dimensional counterparts (e.g.~Schwarzschild or Schwarzschild-AdS spacetimes), computationally expensive mode sums are usually needed. 

The BTZ spacetime is obtained as a quotient of AdS-Rindler spacetime  \cite{hodgkinsonPhysRevD.86.064031,LANGLOIS20062027,Smith_2014} under the identification $\Gamma:\phi \to \phi + 2\pi\sqrt{M}$ (see Supplemental Materials for further details). The line element is
\begin{align}
    \D s^2 &= - f(r) \D t^2 + f(r)^{-1} \D r^2 + r^2 \D \phi^2 ,
\end{align}
where $f(r) = (r^2/l^2 - M)$ and $\sqrt{M}l < r < \infty$, $-\infty< t < \infty$, $\phi \in [0,2\pi)$.  The spacetime has a local Tolman temperature  $T_L = r_H/(2\pi l^2 \sqrt{f(R)})$, where $R$ denotes the radial coordinate and
$r_H = \sqrt{M}l$ is the radial coordinate of the event horizon. 

To construct a quantum field theory on this background, consider an automorphic field $\hat{\phi}(\textsf{x})$, which for the BTZ spacetime is constructed from an ordinary (massless scalar) field $\hat{\psi}$ in (2+1)-dimensional AdS spacetime (AdS$_3$) via the identification
 $\Gamma$, yielding \cite{Langlois:2005if}
\begin{align}\label{eq:field_identif}
    \hat{\phi}(\textsf{x}) := \frac{1}{\sqrt{\mathcal{N}}} \sum_n \eta^n \hat{\psi}(\Gamma^n \textsf{x}) 
\end{align}
where $\textsf{x} =(t,r,\phi) $,
$\mathcal{N} = \sum_n \eta^{2n}$ and $\eta = \pm 1$ (corresponding to untwisted and twisted fields respectively). The Wightman (two-point correlation) function  between 
$\textsf{x},\textsf{x}'$ is  
\begin{align}\label{eq6}
    W_\text{BTZ}^{(D)}(\textsf{x},\textsf{x}') &= \frac{1}{\mathcal{N}} \sum_{n,m} \eta^n \eta^m \langle 0 | \hat{\psi} (\Gamma_D^n \textsf{x} ) \hat{\psi}(\Gamma_D^{m} \textsf{x}' ) | 0 \rangle 
\end{align}
where $\Gamma_{D}^n: (t,r,\phi) \to (t,r, \phi + 2\pi n \sqrt{M_{D}})$
in a  BTZ spacetime with black hole mass $M_{D}$. We have assumed that the field state on the right-hand side of Eq.\ (\ref{eq6}) is the AdS vacuum state; the thermal properties of the black hole  arise from the topological identifications (see \cite{lifschitzPhysRevD.49.1929} for details of constructing the BTZ Wightman function as an image sum of the vacuum AdS Wightman function). Although the normalisation factor $\mathcal{N}$ is formally divergent,   Eq.~\eqref{eq6}, 
 is finite. 

We consider the field quantized on a background arising from superposing BTZ spacetimes with different black hole masses. The black hole--quantum field system can be described in the tensor product Hilbert space $\mathcal{H} = \mathcal{H}_\text{BH} \otimes \mathcal{H}_F$, where we consider the black hole to be (without loss of generality) in a symmetric superposition\footnote{Conceptually, a black hole mass-superposition may be generated by sending a photon prepared in a wavepacket distribution of frequencies into the black hole. This would in general create a black hole in a superposition of energy (i.e.\ mass-energy) eigenstates.}
of two mass states $|M_A\rangle$, $|M_B\rangle$ while the field is in the AdS vacuum $| 0 \rangle$. When coupling a particle detector to the black hole--quantum field system, we will require correlation functions between the fields on the different amplitudes of the superposition;  an analogous procedure (see Supplementary Material)
as in Eq.\ (\ref{eq6}) yields  
\begin{align}
    W_\text{BTZ}^{(AB)}(\textsf{x},\textsf{x}') &= \frac{1}{\mathcal{N}} \sum_{n,m} \eta^n \eta^{m} \langle 0 | \hat{\psi} (\Gamma_A^n \textsf{x}) \hat{\psi} (\Gamma_B^m \textsf{x}' ) | 0 \rangle ,
\end{align}
noting the two different isometries, $\Gamma_A$, $\Gamma_B$, corresponding to the superposed masses, $M_A$, $M_B$, that act on the coordinates of the field operators. 

\textit{Particle detectors}\textemdash To couple matter to 
the quantum 
black hole---field system, we utilize the Unruh-deWitt model, which considers a pointlike two-level system linearly coupled to the field~\cite{RevModPhys.80.787}.

In our case the interaction Hamiltonian reads,
\begin{align}
    \hat{H}_\text{int.} &= \lambda \eta(\tau) \hat{\sigma}(\tau) \sum_{D=A,B} \hat{\phi}(\textsf{x}_{D}) \otimes | M_D \rangle \langle M_D |
\end{align}
where $|\lambda| \ll 1$ is the  coupling constant, $\eta(\tau)$ is a time-dependent switching function, and $\hat{\sigma}(\tau) = ( |e \rangle\langle g | e^{i\Omega\tau} + \text{H.c})$ is the SU(2) ladder operator between the detector's energy eigenstates $|g \rangle$ and $|e \rangle$ with energy gap $\Omega$,  $| M_D \rangle \langle M_D|$ is a projector on the black hole mass, and  $\hat{\phi}(x_D)$ is the field operator for the spacetime associated with black hole mass $M_D$. This interaction means that for each mass $M_D$, the field is identified accordingly, i.e.\ $(t,r,\phi) \to (t,r,\phi + 2\pi n \sqrt{M_D})$. Finally, $\tau$ is a proper time related to the coordinate time $t$ of the BTZ spacetime with mass $M_D$ by $\tau = \gamma_Dt$ where $\gamma_D = \sqrt{f(M_D,R)}$ is a redshift factor evaluated at the radial coordinate of the detector. 
We take the detector to be prepared in its ground state $|g\rangle$ at a fixed radial coordinate $R$, which
for the superposed black hole means
it is in a superposition of proper distances from the  horizon.
The evolution of the system relative to the coordinate time $t$ is described as
\begin{align}\label{eq8}
    | \psi(t_f) \rangle &= e^{-i\hat{H}_{0,S}t_f} \hat{U}(t_i,t_f) e^{i\hat{H}_{0,S}t_i} | \psi(t_i) \rangle 
\end{align}
where $\hat{H}_{0,S}$ is the free Hamiltonian of the whole system, including the evolution of the mass eigenstates of the black hole between the initial and final times $t_i$ and $t_f$. This evolution introduces a relative phase between the constituent states of the superposition. Particular attention must be paid to the definition of $t_i$ and $t_f$. Here, we assume that the black hole--field--detector system is in a tensor product with an external `clock state' whose time coordinate defines the evolution of the whole system. 
For example, for sufficiently large radial coordinate $r$, the fractional difference in proper times of the clock between two spacetimes with mass $M_A$ and $M_B$ respectively is $(M_B - M_A)l^2/2r$. Given the finite precision of any physical clock, such a factorizable clock state can thus be associated with the proper time of a clock at sufficiently large $r$. One should expect a theory of quantum gravity describing spacetime superpositions to admit these kinds of factorizable clock states, as well as clock states that are entangled with the spacetime~\cite{ruiz2017}.

Returning to Eq.\ (\ref{eq8}), we expand the time-evolution operator to leading order in the coupling strength,
\begin{align}
    \hat{U}(t_i,t_f) &= I + \hat{U}^{(1)} + \hat{U}^{(2)} + \mathcal{O}(\lambda^3)
\end{align}
where the first- and second-order terms are
\begin{align}
    \hat{U}^{(1)} &= - i \lambda \int_{t_i}^{t_f} \D \tau \: \hat{H}_\text{int}(\tau) , \vphantom{\iint_{\mathcal{T}}} \\ 
    \hat{U}^{(2)} &= - \lambda^2 \int_{t_i}^{t_f} \D \tau \int_{t_i}^{\tau} \D \tau' \hat{H}_\text{int.}(\tau) \hat{H}_\text{int.}(\tau') .
\end{align}
We evolve the initial state according to Eq.\ (\ref{eq8}) and then look at the final state of the detector conditioned on measuring/projecting the black hole in a mass-superposition basis. This scenario describes a Mach-Zehnder type interference where the interferometric paths are associated with different mass states of the black hole, and yields a detector transition probability conditioned upon the measurement.  If one considers, for example, a measurement in the symmetric-antisymmetric superposition basis, $|\pm \rangle$, then one finds that the ground and excited state probabilities of the detector are 
(see Supplemental Materials)
\begin{align}\label{13}
    P_G^{(\pm)} &= \frac{1}{2}\Big(1 \pm \cos (\Delta E \Delta t) \Big) \Big[ 1 - \frac{\lambda^2}{2} \Big( P_A + P_B \Big) \Big] ,\\ \label{14}
    P_E^{(\pm)} &= \frac{\lambda^2}{4} \Big( P_A + P_B \pm 2 \cos (\Delta E \Delta t )L_{AB} \Big) ,
\end{align}
where $P_{D}$ $(D = A,B)$ is the transition probability of a static detector outside a BTZ black hole with classical mass $M_{D}$, while $L_{AB}$ is a cross-correlation term quantifying the correlations between the 
fields on the spacetime superposition. Explicitly, \begin{align}
    P_D &= \int_{t_i}^{t_f} \D \tau \int_{t_i}^{t_f} \D \tau' \chi(\tau) \bar{\chi}(\tau') W_\text{BTZ}^{(D)}(\textsf{x},\textsf{x}'), \\ 
    L_{AB} &= \int_{t_i}^{t_f} \D \tau \int_{t_i}^{t_f} \D \tau' \chi(\tau) \bar{\chi}(\tau') W_\text{BTZ}^{(AB)} (\textsf{x},\textsf{x}'),
\end{align}
where $\chi(\tau) = \eta(\tau) e^{-i\Omega\tau}$. We have also defined $\Delta E = |\sqrt{M_A} - \sqrt{M_B}|$ as the energy difference between the black hole mass eigenstates and $\Delta t = t_f-t_i$ as the time window over which the interaction is switched on. Note that the probabilities in Eq.\ (\ref{13}) and (\ref{14}), obtained for conditional measurements in the complete basis $| \pm \rangle$, add to unity. 

Calculations utilizing the Unruh-deWitt model typically take $\Delta t \to \infty$; one integrates over the entire history of the detector. In this case, the presence of an oscillating phase, dependent on the mass and time difference, means that this limit is not well-defined\footnote{In the Supplemental Materials, we show how one may utilize the rotating-frame transformation \cite{whaleyPhysRevA.29.1188}, commonly used in quantum optics and atomic physics, to absorb any relative phases into the definitions of the states.}. In our analysis, we consider for the sake of argument arbitrarily chosen values of $t_f$ such that $\sigma \ll \Delta t$ where $\sigma$ is a characteristic timescale for which the detector interacts with the field. Specifically, we consider transition probabilities for Gaussian detector switching functions, $\eta(\tau) = \exp ( - \tau^2/2\sigma^2)$, giving
\begin{align}\label{eq17}
    \frac{P_D}{\sigma} &= \frac{\sqrt{\pi} H_0(0)}{8} - \frac{i }{8\sqrt{\pi}} \text{PV} \int_{-t_f/2l}^{t_f/2l} \frac{\D z \: X_0(2lz)H_0 (2lz)}{\sinh(z)} \non \\
    & + \frac{1}{4\sqrt{2\pi}\sum_n \eta^{2n}} \sum_{n\neq m} \text{Re} \int_0^{t_f/l}  \frac{\D z \: X_0(lz) H_0 (lz) }{\sqrt{ \beta_{nm} - \cosh(z) }}
\end{align}
where $X_0(z)$, $H_0(z)$ are functions of the detector and spacetime parameters derived in the Supplemental Materials. Likewise, the cross-term is given by 
\begin{align}\label{eq18}
    \frac{L_{AB}}{\sigma} &=   \frac{Y_0}{\sum_n \eta^{2n}} \sum_{n,m} \text{Re} \int_0^{t_f/l} \frac{\D z \:  Z_0(lz) Q_0(lz) }{\sqrt{ \alpha_{nm} - \cosh(z)}},
\end{align}
 with $Y_0$, $Z_0(z)$ and $Q_0(z)$ derived in the Supplemental Materials. The constants $\alpha_{nm}$, $\beta_{nm}$ are given by
\begin{align}
    \beta_{nm} &= \frac{1}{\gamma_D^2} \left[  \frac{R^2\cosh(2\pi(n-m)\sqrt{M_D})}{M_Dl^2} - 1 \right] ,
    \\ \label{alphanm}
    \alpha_{nm} &= \frac{1}{\gamma_A\gamma_B} \left[ \frac{R^2\cosh(2\pi (m \sqrt{M_A} - n \sqrt{M_B} ) )}{\sqrt{M_AM_B} l^2} - 1 \right] .
\end{align}
In the Supplemental Materials, we also derive expressions for a compactly supported detector switching.

\begin{figure}[h]
    \centering
    \includegraphics[width=1\linewidth]{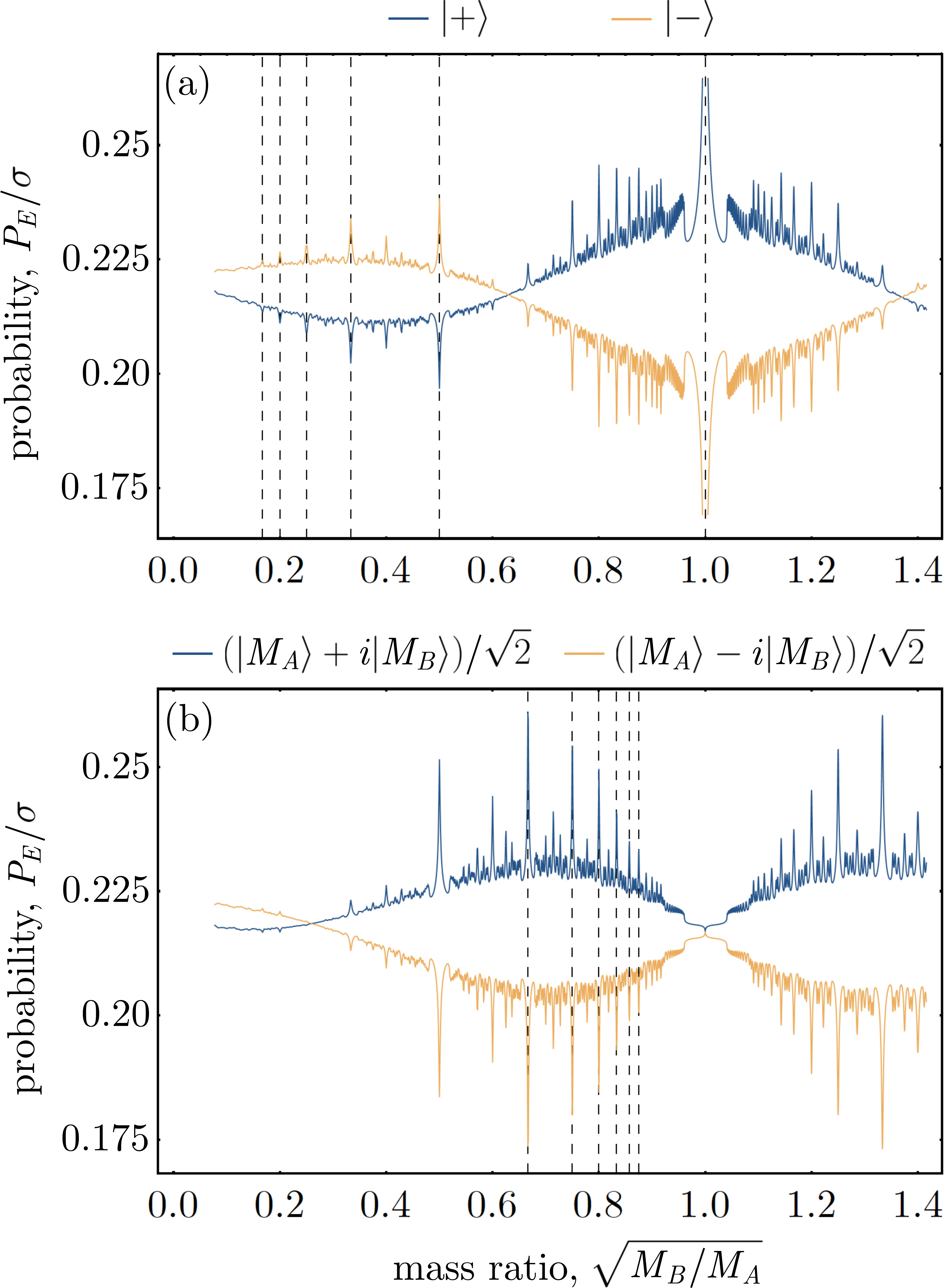}
    \caption{Transition probability of the detector as a function of $\sqrt{M_B/M_A}$. The measurement basis corresponding to the relevant plot is indicated by the legend. In (a), the dashed lines correspond to $\sqrt{M_B/M_A} = 1/n$ where $n = \{1, \hdots 6\}$. In (b), the dashed lines correspond to $\sqrt{M_B/M_A} = (n-1)/n$ where $n = \{3, \hdots 8\}$. Moreover, the oscillating cross term in (b) is $\pi/2$ out-of-phase with that for the black hole measured in the (anti)symmetric basis. In all plots we have also used $l/\sigma = 5$, $R/ \sigma = 25$, $t_f = 5\sigma$ and $M_Al^2 = 2$.}
    \label{fig:plot1}
\end{figure}

\textit{Results}\textemdash 
We can now analyze the response of the detector outside the mass-superposed black hole. In Fig.\ \ref{fig:plot1}, we have plotted the response of the detector as a function of the mass ratio of the black hole superposition, $\sqrt{M_B/M_A}$.
Fig.\ \ref{fig:plot1}(a) shows the conditional transition probabilities for measurements in the symmetric and antisymmetric superposition basis, while Fig.\ \ref{fig:plot1}(b) considers measurements in the $|i\pm\rangle = ( |M_A \rangle \pm i | M_B\rangle)/\sqrt{2}$ basis. The most interesting features are the resonant peaks that occur at rational values of the square root ratio of the superposed masses. In Fig.\ \ref{fig:plot1}(a), we have denoted some of these ratios with dashed lines. We ascribe this behaviour to a constructive interference  between the field modes associated with 
topologically closed AdS spacetimes, yielding resonances in the detector response at integer values of $\sqrt{M_B/M_A}$. While we have 
shown the transition probability for $M_B<M_A$, these resonances also occur for $M_B>M_A$. The effect also occurs for other values of the square root mass ratio -- in Fig.~\ref{fig:plot1}(b), we have highlighted the ratios $\sqrt{M_B/M_A} = (n-1)/n$ for $n \in \{ 3, \hdots 8\}$. To understand this, note that in Eq.\ (\ref{alphanm}), the argument of $\cosh(x)$ vanishes when $m\sqrt{M_A} = n \sqrt{M_B}$. These `coincidences' occur for terms in the image sum when the ratio $\sqrt{M_A/M_B}$ is a rational number, which includes ratios allowed by Bekenstein's quantum black hole conjecture. Specifically, the allowed mass values of the BTZ black hole, assuming the Bohr-Sommerfeld quantization scheme for the horizon radius, are given by \cite{Kwon_2010}:
\begin{align}\label{eq19}
    r_H &= \sqrt{M}l = n , \qquad n = 1, 2, \hdots 
\end{align}
While our construction does not require that the superposed masses are quantized in integer values, the form of Eq.\ (\ref{alphanm}) explains the origin of the signatory resonances. The detector responds uniquely to black hole mass superpositions with mass ratios corresponding to the masses predicted by Bekenstein's conjecture. This result provides independent support for Bekenstein's conjecture, demonstrating how the detector's excitation probability can reveal a genuinely quantum-gravitational property of a quantum black hole.

Note also the low-frequency oscillation in the transition probability as $\sqrt{M_B/M_A}$ is varied. This behaviour arises from the 
evolution of the black hole superposition, which depends on the energy difference between the  superposed states. As $\sqrt{M_B/M_A} \to 1$, the transition probability approaches that of a single black hole for a measurement in $|+\rangle$, whereas it vanishes for $|- \rangle$. 
For measurements in $|i\pm \rangle$, the transition probability approaches half of that of a detector situated outside the black hole with a classical mass value $M_A = M_B$. In Fig.~\ref{fig:plot1}(a), the amplitude of the oscillation decreases in the limit of $\sqrt{M_B/M_A} \ll 1$ or $\gg 1$ due to the decay of the correlation term $L_{AB}$. 

\textit{Metric for spacetime superpositions}\textemdash Our approach has allowed us to study the quantum effects of a spacetime (black hole) superposition through the calculation of a physical observable -- the transition probability of a 
detector coupling to the field. This observable is constructed from the two-point correlations of the field, which remarkably can be done also on a non-classical spacetime, such as that of a BTZ black hole in  superposition of masses. Furthermore, it is possible to write down the associated spacetime metric using these correlators. We first review a related argument relating metric and field correlators by Saravani et.\ al.\ and Kempf \cite{Saravani_2016,Kempf:2021xlw}. Recall that in general relativity, spacetime is described by the pair $(M,\sigma)$ where $M$ is a differentiable manifold and $\sigma(\textsf{x},\textsf{x}')$ is the Synge world function (or the geodesic distance) for two points $(\textsf{x},\textsf{x}')$~\cite{wald2010general,synge1960relativity,fulling1989aspects,birrell1984quantum}. The latter contains all the information about a spacetime, since it allows one to reconstruct the metric~\cite{wald2010general,synge1960relativity,fulling1989aspects,birrell1984quantum}:
\begin{align}
    g_{\mu\nu}(x) &= - \lim_{x\to x'} \frac{\p}{\p x^\mu} \frac{\p}{\p x'^{\nu}} \sigma(\textsf{x},\textsf{x}') .
\end{align}
Since the quantum correlations of a field decay with the magnitude of the distance between spacetime points, these studies replaced the notion of proper distance with that of an appropriate measure of  such correlations. In \cite{Kempf:2021xlw}, the Feynman propagator is used, but other objects like the Wightman function are also suitable. Using the Wightman function, the spacetime metric can be expressed as 
\begin{align}\label{eq3}
    g_{\mu\nu} &= {\Upsilon(d)}\lim_{x\to x'} \frac{\p}{\p x^\mu} \frac{\p}{\p x'^{\nu}} W(\textsf{x},\textsf{x}')^{\frac{2}{d-2}}
\end{align}
where  $\Upsilon(d) = - (1/2)( \Gamma(d/2 - 1 )/(4\pi^{d/2}))^{\frac{2}{d-2}}$ and $d>2$ is the spacetime dimension\footnote{A unique expression exists for $d = 2$}. In this context, building a spacetime in superposition occurs at the level of the field operator. By taking $\hat{\phi}(x) \to \hat{\phi}(\textbf{x}) = \sum_D f_D \hat{\phi}(x_D)$  (where $\textbf{x} = \{ x_D \}$, $\sum_D |f_D|^2 = 1$ and the relative phases between $f_D$ are determined by the state in which the black hole is measured), the Wightman function becomes a sum over all two-point correlators between the fields $\hat{\phi}(x_D)$, $\hat{\phi}(x_{D'}')$, defined with respect to the coordinates of the spacetime states in superposition.  Equation (\ref{eq3}) is then modified as follows, yielding a  conditional metric describing a superposition of spacetimes:
\begin{align}\label{21}
    g_{\mu\nu} &= \Lambda(d) \lim_{x\to x' } \frac{\p}{\p x^\mu} \frac{\p}{\p x'^{\nu}} \sum_{D,D'} f_D f_{D'}^\star W(x_D,x'_{D'})^{\frac{2}{d-2}}.
\end{align}
Equation (\ref{21}) involves correlations between the field operators parametrized with coordinates covering black hole spacetimes associated with different masses; it represents a `conditional metric'  effectively seen by a detector in the quantum superposition of spacetimes.

\textit{Conclusion}\textemdash In this Letter, we have derived genuine quantum-gravitational phenomena via a simple particle detector model in a spacetime with quantum degrees of freedom. In particular, we have developed an operational framework for analyzing `superpositions of spacetimes' via matter (represented by an Unruh-deWitt detector) coupled to a quantum field. We applied this approach to study the effects produced by a BTZ black hole in a superposition of masses. The response of a detector in this spacetime features `cross-correlations' between the different spacetime amplitudes. 

The response of the detector is sensitive to the mass ratio of the superposed black hole, in particular exhibiting signatory peaks at rational values of $\sqrt{M_B/M_A}$. This   description
of a quantum-gravitational effect is the first of its kind, and constitutes a new method for investigating effects implied by Bekenstein's conjecture about the quantization of a black hole's mass \cite{BEKENSTEIN19957}. We also presented a description of a conditional spacetime metric arising from such scenarios which is constructed using field correlations as a measure of distance \cite{Saravani_2016,Kempf:2021xlw}. Finally, while we have studied the analytically tractable 
(2+1)-dimensional black hole, there is no in-principle obstruction from applying our approach to more general cases in (3+1)-dimensions, which are expected to provide additional insight into this important problem.

The scheme presented here opens a path to further understanding important quantum gravity ideas like quantum black holes, spacetime superpositions, and quantum-gravitational causal structures. Such notions are considered crucial for a complete description of quantum gravity; however they have generally only been considered using top-down perspectives in the literature. Our approach allows for the calculation of observables, such as the response of a first-quantized system to quantum fields, without the need for a complete framework of quantizing gravity. By connecting the notion of quantum field correlations with the superposed spacetime metric, we have also opened up the possibility of studying the dynamics of spacetime superpositions and the effects they may induce on systems such as low-energy particles, which continues to be a topic of growing interest. 

\textit{Acknowledgements}\textemdash
This work was supported in part by the Natural Sciences and Engineering Research Council of Canada. M.Z.\ acknowledges support from Australian Research Council (ARC) Future Fellowship, grant FT210100675, and ARC Centre of Excellence EQUS CE170100009. The University of Queensland (UQ) acknowledges the Traditional Owners and their custodianship of the lands on which UQ operates.

\normalem

\bibliography{main}

\end{document}